\documentclass[mathleft]{an}
\usepackage{graphicx}
\usepackage{times}
\begin{document}

\Pagespan{1057}{1060}
\Yearpublication{2008}%
\Yearsubmission{2008}%
\Month{?}%
\Volume{329}%
\Issue{9/10}%
\DOI{10.1002/asna.200811090}%

\title{Intracluster Planetary Nebulae in the Hydra I Cluster\thanks{Based on data collected at ESO, Chile (ESO program 076.B-0641(B))} }

\author{G. Ventimiglia\inst{1,2}\fnmsep\thanks{Corresponding author:
  \email{gventimi@eso.org}\newline}
\ M. Arnaboldi\inst{1}
          \and
          O. Gerhard\inst{2}
}
\titlerunning{Intracluster Planetary Nebulae in the Hydra I Cluster}
\authorrunning{G. Ventimiglia, \ M. Arnaboldi \& O. Gerhard}
   \institute{ESO, Karl-Schwarzschild-Strasse 2, D-85748 Garching bei Muenchen\\
         \and
            Max Planck Institute fuer Extraterrestrische Physics,
              Giesenbachstrasse, D-85741 Garching bei Muenchen\\ 
             }

\received{2008 Sep 18}
\accepted{2008 Sep 22}
\publonline{2008 Nov 20}


\abstract{%
  Using the Multislit Imaging Spectroscopy (MSIS) technique at the
  FORS2 spectrograph on VLT-UT1, we have identified 60 Intracluster
  Planetary Nebula (ICPN) candidates associated with the Intracluster
  Light (ICL) in the central region of the Hydra I cluster. Hydra I is
  a medium compact, relatively near ($\sim50$ Mpc), rich cluster in
  the southern hemisphere. Here we describe the criteria used to
  select emission sources and present the evidence for these PN
  candidates to be associated with the ICL in the Hydra I cluster. We
  also show, using the luminosity-specific planetary nebulae number,
  the $\alpha$ parameter, that the expected number of PNs associated
  with the stellar population of the central cD galaxy NGC 3311 in the
  cluster is close to the number of PNs detected.}

\maketitle

\section{Introduction}
Diffuse intracluster light (ICL) has been observed both in nearby and
intermediate redshift clusters (Mihos et al. 2005, Feldmeier et
al. 2004). Observations show that the diffuse light in galaxy clusters
amounts to $\sim10\%$ of the total light emitted by the cluster
galaxies (Aguerri et al. 2005; Zibetti et al. 2005).

Cosmological simulations of structure formation predict intracluster
stars to be lost from galaxies in interactions during the assembly of
galaxy clusters (Na\-politano et al. 2003; Murante et al. 2004;
Willman et al. 2004; Sommer-Larsen, Romeo \& Portinari
2005). Simulations suggest that most of the diffuse light in galaxy
cluster cores originates from mergers that lead to the formation of
the brightest cluster galaxy and of the other massive galaxies, while
the tidal stripping mechanism dominates the formation of the ICL at
larger radii (Murante et al. 2007).

The study of the properties of this diffuse component has then an
important role in understanding the mechanisms relevant for the
evolution of galaxies in high density environments, and the formation
history, dynamics and merging status of clusters.

Wide field imaging has already shown the morphological complexity of
the ICL: studies of the Virgo and Coma cluster ICL have shown that it
is characterized by tidal features like streamers and extended galaxy
halos (Adami et al. 2005; Mihos et al. 2005).  Due to its
intrinsically low surface brightness, $\mu_V>28.5$ mag arcsec$^{-2}$
(Feldmeier et al. 2004; Mihos et al. 2005), the kinematics of ICL can
only be studied using the \textit{Intracluster Planetary Nebulae}
(ICPNs) associated with this stellar component (Arnaboldi et al. 2004,
Gerhard et al. 2007, Doherty et al. 2008). ICPNs are relatively easy
to detect because their spectra are characterized by two strong
emission lines: [OIII]$\lambda4959$\AA\ and [OIII]$\lambda5007$\AA,
with relative flux ratio $1:3$.

The goal of our project is to measure the velocity distribution of the
PNs associated with the diffuse light in the central region of the
nearby Hydra I cluster (Abell 1060, $D\sim50$ Mpc, $z\sim0.0126$), a
medium compact, non-cooling flow, rich cluster in the southern
hemisphere. This will be presented in a forthcoming paper. Here we
present the selection criteria and the evidence for the detection of
PNs associated with the ICL in Hydra I.

At the distance of the Hydra I cluster, the flux of the O[III]
$\lambda5007$\AA\ emission line of a PN is less than $8\times
10^{-18}$ erg s$^{-1}$cm$^{-2}$, therefore we need to reduce
substantially the noi\-se from the night sky in order to detect these
lines. This is possible by using the Multislit Imaging Spectroscopy
technique (MSIS: Gerhard et al. 2005, Arnaboldi et al. 2007).  In what
follows we define $m_{5007}=-2.5\log{F_{5007}}-13.74$ (the Jacoby
formula, Jacoby et al. 1989), where $F_{5007}$ is the integrated flux
in the [OIII] $\lambda5007$\AA\ emission line, and we assume a
distance of 50 Mpc for the Hydra I cluster implying $1"\sim0.24$ Kpc.

In the next section we present the MSIS observations carried out with
FORS2 on UT1. In Section 3 we summarize the data reduction.  We
present the adopted selection criteria to identify the emission
sources and the evidence for PNs associated with the ICL in the Hydra
I cluster in Section 4. In Section 5 we show that the number density
of PN candidates detected is consistent with that expected for the
stellar population associated with the central cD galaxy of the
cluster, NGC 3311. In Section 6 we summarize our results.

\section{Observations}

\subsection{The Multislit Imaging Spectroscopy technique}
The MSIS technique consists of the combined use of a mask of parallel
slits, a narrowband filter centered around the redshifted [OIII]
$\lambda5007$\AA\ line, and a dispersing element.  It is a blind
search technique and allows one to obtain spectra of all PNs (and
other emission line objects) that happen to lie behind the mask
slits. Because the [OIII] emission lines from PNs are only $\sim 30$
km/s wide (Arnaboldi et al. 2008), when dispersed at spectral
resolution $R=\lambda/\Delta\lambda\sim6000$, their entire flux falls
into a small number of pixels in the two-dimensional spectrum,
determined by the slit width and seeing. By dispersing the sky noise
on many pixels, the technique allows measurement of very faint
fluxes. We can detect PNs with magnitudes $\sim 1.2$ mag below the
bright cut-off of the Planetary Nebulae Luminosity Function (PNLF;
Ciardullo et al. 1998) and their positions and radial velocities can
be measured at the same time.

\subsection{Observational set up}
Data were collected in visitor mode during 2006 March 26-28, using the
FORS2 spectrograph on UT1.  The observed area covers the central
region of the Hydra I cluster, around NGC 3311, at
$\alpha=$ 10h36m42.8s, $\delta$= -27d31m42s (J2000).

The FORS2 field of view (FOV) is $6'.8\times6'.8$ wide, corresponding
to $\sim10000$ kpc$^2$, and is imaged onto a mosaic of two CCDs,
rebinned $2\times2$ in the readout.  We used two narrow band filters,
one centered at $\lambda=5045$\AA\ and a second one at
$\lambda=5095$\AA, both with a FWHM of $60$\AA. We cover, in this
way, the whole range of cluster line-of-sight (LOS) velocities and,
for fast, $v_{\rm LOS}\geq 4000$ km s$^{-1}$, and bright PNs,
$m_{5007}\leq 29.3$, the [OIII] $\lambda5007$\AA\ emission line falls
in the red filter and the [OIII] $\lambda4959$\AA\ line in the blue
filter.  Spectra were obtained with the GRIS-600B grism, which has a
spectral resolution of 0.75 \AA/pixel at 5075 \AA. The MSIS mask is
made of $24\times21$ slits, each of them $0".8$ wide and $17".5$
long. Each slit is projected along the dispersion axis onto $\sim 40$
rebinned pixels. The effective area imaged by the slits is $\sim 7056$
arcsec$^2$, that is $\sim4.5\%$ of the whole FORS2 FOV. In order to
cover the whole field the MSIS mask was stepped 15 times on the sky to
cover the region between two adjacent slits.  For each mask position 3
exposures, of $1200$ sec each, were taken, ensuring a proper cosmic
ray subtraction.  The seeing during the three observing nights was in
the range from $0".6$ to $1".5$.

Arclamp calibration frames were acquired for the wavelength
calibration, as well as flats and bias images.  The flux calibration
was done using long slit observations of the standard star LTT7379
with narrow band filter plus Grism.

\section{Data reduction}
Data reduction was carried out following the procedure described in
Arnaboldi et al. (2007).  After bias subtraction, the images were
properly co-added and the continuum light from the two main galaxies
was subtracted, with an fmedian filtering using a window of
$19\times35$ pixels.

\begin{figure} \centering
     \includegraphics[width=7.5cm]{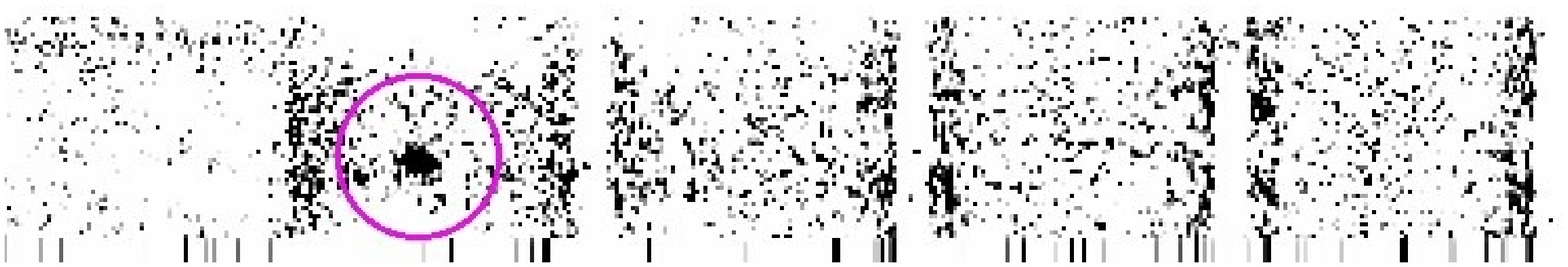}
     \includegraphics[height=4cm,width=6cm]{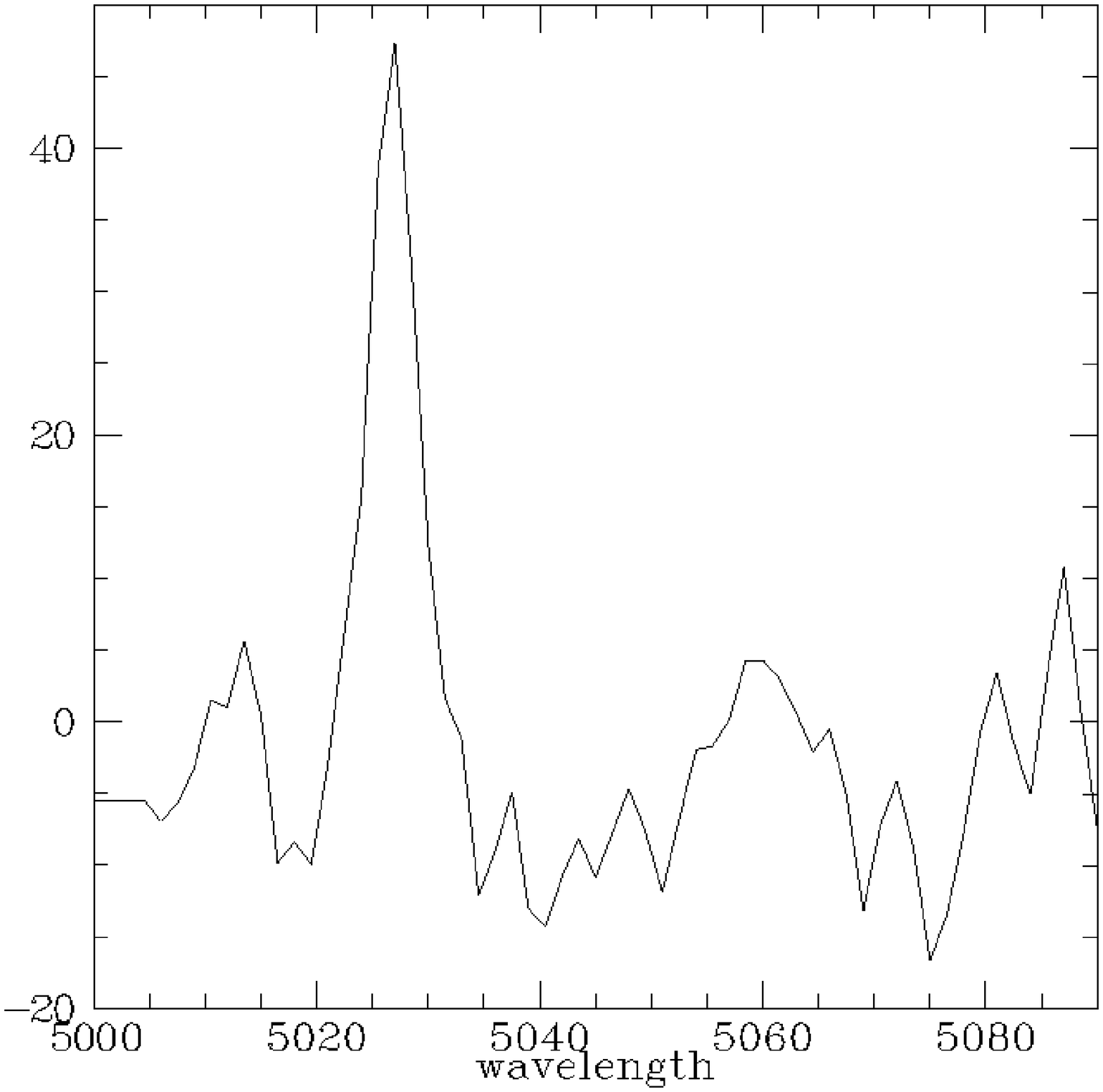}
     \caption{\textit{Upper panel}: 2-D spectrum of a PN
     candidate. The dispersion axis is along the vertical
     direction. The emission line falls onto about 6 pixels both in
     the spatial and spectral direction. \textit{Lower panel}: 1-D
     spectrum of the same PN candidate. The emission line has the same
     FWHM as the arc lamp lines, showing that the observed emission
     line is unresolved in wavelength.  } \label{2dspectra}
     \end{figure}

Then we extracted and rectified the 2D spectra (Fig.~\ref{2dspectra})
of the emission sources and performed the wavelength and flux
calibration. The total number of emission sources detected is 95. On
the basis of the flux calibration the $1\sigma$ limit on the continuum
is $\sim7\times10^{-20}$ erg cm$^{-2}$ s$^{-1}$ \AA$^{-1}$.

A first classification of the emission line objects can be carried out
according to the following criteria:
 \begin{itemize}
     \item PN candidates: unresolved emission line objects, both in
     space and in wavelength, with no continuum;
     \item background objects candidates: continuum sources with
     unresolved or resolved emission line.
 \end{itemize}
Of the 95 emission lines sources identified, 60 were classified as
possible PN candidates, the remaining sources as background object
candidates. Monochromatic point like emissions appear in the final
images as unresolved sources with a width of $\sim6$ pixels both in
the spatial and in the wavelength direction. The final spectral
resolution is $\sim4.5$\AA\, or $140$ km s$^{-1}$.  Magnitudes for
the PN candidates were computed using the Jacoby formula.

\section{Selection Criteria for the PN population in the Hydra cluster: 
the wavelength-magnitude plane}
\begin{figure}
     \centering
     \includegraphics[width=7cm]{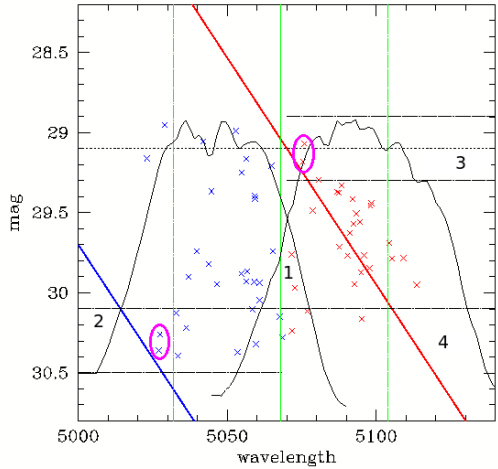}
        \caption{The plot shows the wavelength vs magnitude space in
        which we describe the physical properties of a PN population
        at 50 Mpc and bound to the Hydra I cluster. Blue and red
        crosses are the PN candidates detected in the MSIS images. The
        various lines and the characteristics of the 4 regions
        identified in the plot are explained in the text.}
           \label{lambda_mag}
     \end{figure}

We now describe the physical properties of PNs belonging to the Hydra
I ICL in a two dimensional wavelength-magnitude space, as shown in
Fig.\ref{lambda_mag}.

A PN population is characterized by a bright cut-off of the PNLF,
which according to Ciardullo et al. (1998) has absolute magnitude
of the [OIII] $\lambda5007$\AA\ emission line of $M^*=-4.48$. At the
Hydra I cluster distance ( $m-M=33.49$), this corresponds to an
apparent magnitude of $m_{5007}=29.1$ (plotted as the dotted
horizontal line in Fig.\ref{lambda_mag}). The apparent magnitude for
the [OIII] $\lambda4959$\AA\ emission line is 1.2 mag fainter.

The red inclined line in Fig.\ref{lambda_mag} shows the apparent
magnitude of a PN at the PNLF bright cutoff, as a function of the
Hubble flow distance, and the blue line shows the same dependence for
the [OIII] $\lambda4959$\AA\ emission line. If the line-of-sight
velocity corresponded only to the Hubble flow, then the magnitude of
the two emission lines, [OIII] $\lambda4959$\AA\ and [OIII]
$\lambda5007$\AA\ of a PN fainter than the bright cut-off would fall
on the left side of the blue and red line, respectively.

The systemic velocity of the Hydra I cluster $v_{\rm Hydra\,I}=3683$
km s$^{-1}$ and its velocity dispersion $\sigma_{\rm Hydra\,I}=724$ km
s$^{-1}$ (Christlein \& Zabludoff 2003) identify a wavelength range in
which [OIII] emissions from objects orbiting in the Hydra I cluster
potential can be observed. At the systemic velocity of the cluster,
the [OIII] emission lines are redshifted to $\lambda=5068$\AA\ and
$\lambda=5020$\AA, respectively.  In Fig.\ref{lambda_mag} the central
green line is at $\lambda=5068$\AA\ where the [OIII]
$\lambda5007$\AA\ line is redshifted to the cluster systemic velocity.
The other two vertical green lines are at the wavelengths bounding the
$\pm3\sigma$ velocity range, respectively at $\lambda=5032$\AA\ and
$\lambda=5104$\AA.

The black continuous lines in Fig.\ref{lambda_mag} show the measured
profiles of the narrow band filters, $T(\lambda)$, normalized so that
the maximum transmission is at the PNLF bright cutoff.  They are
centered respectively at $\lambda=5045$\AA\ and $\lambda=5095$\AA\ and
have a FWHM of $\sim60$\AA.

Considering the magnitude limit and wavelength range spe\-cified, we
can identify different areas in this wavelength-magnitude space,
separated by black horizontal lines:
  \begin{itemize}
     \item \textbf{Region \#1} According to their velocity and
     magnitude all emission lines are [OIII] $\lambda5007$\AA.
     \item \textbf{Region \#2} It is the region where, considering the
     flux and wavelength, we can find both faint [OIII]
     $\lambda5007$\AA\ or [OIII] $\lambda4959$\AA\ emission lines.
     \item \textbf{Region \#3} Here we have high flux and red
     wavelengths so that we can see the bright [OIII]
     $\lambda5007$\AA\ emissions. For such emissions we expect to
     detect the corresponding [OIII] $\lambda4959$\AA\ in region \#2.
     \item \textbf{Region \#4} Here the emissions are most likely
     [OIII] $\lambda5007$\AA. In principle, we may find both
     [OIII]$\lambda\-4959$\AA\ and [OIII]$\lambda5007$\AA; however, if
     an emission were identified as [OIII] $\lambda4959$\AA\, its LOS
     velocity would be about $8600$ km s$^{-1}$ which is more than
     $6\sigma$ away from the cluster systemic velocity. This PN would
     then not be bound to Hydra I and its velocity driven by the
     Hubble flow. Then its [OIII] $\lambda4959$\AA\ magnitude should
     fall on the left side of the blue line. Therefore [OIII]
     $\lambda4959$\AA\ emission lines in this region are ruled out.
  \end{itemize}

We can now populate the wavelength-magnitude plane with the PN
candidates detected in the MSIS spectra. This gives us important
information to validate the PNs catalogue. In the plot the blue and
red crosses are the PN candidates detected in the wavelength ranges
covered by the blue and red filters respectively.

The first result is that the fluxes of these candidates are all
consistent with those of PNs at the distance of the Hydra I cluster,
in the range from $1.7\times10^{-18}$ to $8.4\times10^{-18}$ erg
s$^{-1}$cm$^{-2}$, i.e. $30.7\;>\;m_{5007}\;>\;28.9$.

The second is that for the two bright [OIII] $\lambda5007$\AA\
emission sources identified in region \#3, the corresponding [OIII]
$\lambda4959$\AA\ have been identified in region \#2: these objects
are encircled in magenta. All others emission sources in the plot are
[OIII] $\lambda5007$\AA\ if in the cluster.

Moreover, we know that the $1\sigma$ continuum upper limit flux in
both filters is $\sim7\times10^{-20}$ erg cm$^{-2}$
s$^{-1}$\AA$^{-1}$. Considering that the compact H{\small II} regions
detected in the Virgo cluster (Gerhard et al. 2002) have a V-band
continuum flux of $\sim8.2\times10^{-19}$erg cm$^{-2}$ s$^{-1}$
\AA$^{-1}$, corresponding to $\sim9.6\times10^{-20}$erg
cm$^{-2}$s$^{-1}$\AA$^{-1}$ at the distance of the Hydra I cluster,
this allows us to rule out from our sample compact H{\small II}
regions such as or brighter than those observed in Virgo.  Also, from
the continuum upper limit flux we calculate that the Equivalent Width
(EW) of the most luminous candidates is $EW>90$\AA. This value is
larger than those of [OII] emitting background galaxies at $z=0.37$,
which have $EW_{[OII]}<50$ (Hogg et al. 1998). Therefore, we can rule
out all contaminations to our sample except a few background
Ly$\alpha$ galaxies.

In summary, from analyzing the PN candidates in the
wavelength-magnitude plane we learn that: \begin{itemize}
     \item fluxes are consistent with PNs at the distance of the Hydra
     I cluster; the only possible contaminants are Ly$\alpha$ emitters
     at high redshift;
     \item objects in magenta circles can reliably be classified as
     PNs because we are able to see both their [OIII]
     $\lambda5007$\AA\ and [OIII] $\lambda4959$\AA\ emission lines;
     \item for all other PNs we have detected the [OIII]
     $\lambda5007$\AA\ at the LOS velocity expected for objects bound
     to the Hydra I cluster.
  \end{itemize} 

\section{The PN population associated with the stellar population of the Hydra I cluster}
We now evaluate the expected number of PNs associated with the total
stellar luminosity of NGC 3311 and we test whether the number of
detected candidates is consistent with that expected given the
sampled light, the flux limit and the average color of the stellar
population.  The luminosity-specific
planetary nebulae number, the so called $\alpha$ parameter, 
specifies the number of PNs associated with the amount of light
emitted by the stellar population of a galaxy (Jacoby et
al. 1980). According to Buzzoni, Arnaboldi, \& Corradi (2006) this is
defined as $\alpha=N_{PN}/L_B$, where $N_{PN}$ is the number of PNs of
a galaxy and $L_B$ is its total stellar luminosity in the B-band. As
shown in Buzzoni et al. (2006), the $\alpha$ parameter is correlated
with the $(B-V)$ colour index of the galaxies.

To compare the theoretical number of PNs expected for NGC 3311 in
Hydra I with the number detected in the MSIS images we have gone
through the following steps. We took $(B-V)=1.04$ for NGC 3311 from
Prugniel \& Heraudeau (1998) and the colour excess $E(B-V)=0.079$ from
the NED database to determine the $(B-V)_0=0.961$ colour corrected for
extinction. From this value we obtained $\alpha=1.5\times10^{-7}$
using Table \#6 in Buzzoni et al. (2006). To calculate the
predicted number of PNs in the MSIS images a completeness correction
is necessary. Since with our technique we detect only PNs down to 1.2
magnitude from the PNLF bright cut-off, the number we need is
$\alpha_{1.2}$. This is simply a fraction of the total
luminosity-specific PN number: $\alpha_{1.2}=\alpha/30=5\times10^{-9}$
(Buzzoni et al. 2006).  The total B-band luminosity of the diffuse
light from NGC 3311's envelope in the FORS2 field, excluding the central
$34"$ where the galaxy is brighter than the sky, is
$L_B=6.8\times10^{10}L_{\odot}$ ($M_B=-21.7$ from Vaesterberg,
Lindblad \& Jorsater 1991). Then the total number of PNs expected in
the FOV is: $N_{PN}=\alpha_{1.2}/L_B\sim114$.  Because each MSIS mask
covers an area of about 4.5\% of the total FORS2 area, we thus expect
to see $114\times0.045\sim 5$ PNs per mask, i.e. $\sim75$ PNs in the
total MSIS image sample. This is consistent within $1.5\sigma$ with
the actual number of PN candidates, 60.

\section{Conclusions}
In this work, we have presented MSIS observations of PN candidates
associated with the diffuse light around the central cD galaxy NGC
3311 in the Hydra I cluster. We have discussed criteria used for
selecting these emission sources and have analyzed their properties in
the wavelength-magni- tude plane. In total, we have identified 60 PN
candidates around NGC 3311, consistent with the number of PNs expected
from the diffuse envelope of this galaxy.

In the next steps of our analysis we will focus on the kinematics of
the diffuse stellar population in the halo of NGC 3311. We will study
the histogram of the LOS velocity distribution of the PNs candidates
to obtain information about possible substructures in the ICL and the
merging status of the cluster. Together with the X-ray temperature
profile (Yamasaki et al. 2002) and the surface brightness profiles of
the stellar light this will give us constraints on the
orbit distribution in the halo of this cD galaxy.

\acknowledgements
G. Ventimiglia is supported by an ESO studentship.

\end{document}